%% file: paper.tex
\documentclass[nonacm,sigconf,10pt,screen]{acmart}

\renewcommand\footnotetextcopyrightpermission[1]{} 
\AtBeginDocument{%
  \providecommand\BibTeX{{%
    \normalfont B\kern-0.5em{\scshape i\kern-0.25em b}\kern-0.8em\TeX}}}

\setcopyright{none}
\copyrightyear{}
\acmYear{}
\acmConference[]{}

\usepackage{times}
\usepackage{microtype}
\usepackage{setspace}
\usepackage{etex}
\usepackage{booktabs} 
\usepackage{xcolor}
\usepackage{balance}
\usepackage{url}
\usepackage{graphicx}
\usepackage{array}
\usepackage{multirow}
\usepackage[ruled,vlined,linesnumbered]{algorithm2e}
\usepackage{verbatim}
\usepackage{comment}
\usepackage[normalem]{ulem}
\usepackage{enumitem}
\usepackage{booktabs}
\usepackage{endnotes}
\usepackage{subfig}
\usepackage{enumitem}
\usepackage{paralist} 

\def\Snospace~{\S{}}
\def\Fnospace~{\mbox{Fig.\hspace{0.25em}}}
\def\Tnospace~{\mbox{Tab.\hspace{0.25em}}}
\def\Enospace~{\mbox{Equation\hspace{0.25em}}}

\definecolor{lightgray}{rgb}{.85,.85,.85}  
\definecolor{orange}{RGB}{255,127,0}
\newcommand{\mynote}[1]{}

\newcommand{\eat}[1]{}

\def\compactify{\itemsep=0pt \topsep=0pt \partopsep=0pt \parsep=0pt}
\let\latexusecounter=\usecounter

\def\CPP{{C\nolinebreak[4]\hspace{-.05em}\raisebox{.4ex}{\tiny\bf ++}}}

\begin{document}

\title[MLOS: An Infrastructure for Automated Software Performance Engineering]{MLOS: An Infrastructure for Automated\\ Software Performance Engineering}


\author{
Carlo~Curino,
Neha~Godwal,
Brian~Kroth,
Sergiy~Kuryata,
Greg~Lapinski,
Siqi~Liu,
Slava~Oks,
Olga~Poppe,
Adam~Smiechowski,
Ed~Thayer,
Markus~Weimer,
Yiwen~Zhu
}
\email{first.last@microsoft.com}

\renewcommand{\shortauthors}{C. Curino, et al.}

\input{00_abstract.tex}





\renewcommand{\LARGE}{\Large}

\maketitle

\captionsetup[figure]{font=small}

\input{10_intro}
\input{30_architecture}

\input{50_evaluation}
\input{60_related}
\input{70_conclusion}

\balance
\bibliographystyle{abbrv}
\bibliography{references}

\end{document}

%% file: 00_abstract.tex

\begin{abstract}
Developing modern systems software is a complex task that combines business logic programming and Software Performance Engineering (SPE). The later is an experimental and labor-intensive activity focused on optimizing the system for a given hardware, software, and workload (hw/sw/wl) context.

Today's SPE is performed during build/release phases by specialized teams, and cursed by:
1) lack of standardized and automated tools,
2) significant repeated work as hw/sw/wl context changes,
3) fragility induced by a ``one-size-fit-all'' tuning (where improvements on one workload or component may impact others).
The net result: despite costly investments, system software is often outside its optimal operating point---anecdotally leaving 30 to 40\% of performance on the table.

The recent developments in Data Science (DS) hints at an opportunity: combining DS tooling and methodologies with a new developer experience to transform the practice of SPE. In this paper we present: \emph{MLOS, an ML-powered infrastructure and methodology to democratize and automate Software Performance Engineering. MLOS enables continuous, instance-level, robust, and trackable systems optimization.} MLOS is being developed and employed within Microsoft to optimize SQL Server performance. Early results indicated that component-level optimizations can lead to 20\%-90\% improvements when custom-tuning for a specific hw/sw/wl, hinting at a significant opportunity. However, several research challenges remain that will require community involvement.
To this end, we are in the process of open-sourcing the MLOS core infrastructure, and we are engaging with academic institutions to create an educational program around Software 2.0 and MLOS ideas.

\end{abstract}

%% file: 10_intro.tex

\section{Introduction}

Software Performance Engineering (SPE) is hard because performance depends on the software and hardware environment where the system is run \emph{and} what workload it is processing.
This situation is even more difficult, since these influences are not static in time. In response to these challenges, real-world SPE has developed as an amorphous collection of experiments and tribal best practices aimed at tuning parameters and policies to satisfy non-functional performance requirements (e.g., latency, throughput, utilization, etc.).
Consider as an example the design and tuning of a hash table. Implementing a working in-memory hash table is relatively simple.
But tuning it requires a deep understanding of the workload, complex hardware capabilities and interactions, and local as well as global performance/resource trade-offs of the entire system---as shown in our experiments in \S\ref{sec:eval}. 

Our key observation is that: while {\em business logic development} is a necessary and high-value activity, {\em manual tuning of parameters/heuristics} is an often low-yield (yet important) activity for engineers.
The key reasons for the current situation are: 1) lack of integrated tools to explore a large parameter space by running, and tracking experiments, 2) a need to continuously re-tune systems as hardware/software/workload (hw/sw/wl) context evolve, and 3) the manual nature of the process, that forces one-size-fits-all configurations.

\begin{figure*}[h!]
    \centering
    \vspace{-2mm}
    \includegraphics[width=0.9\textwidth]{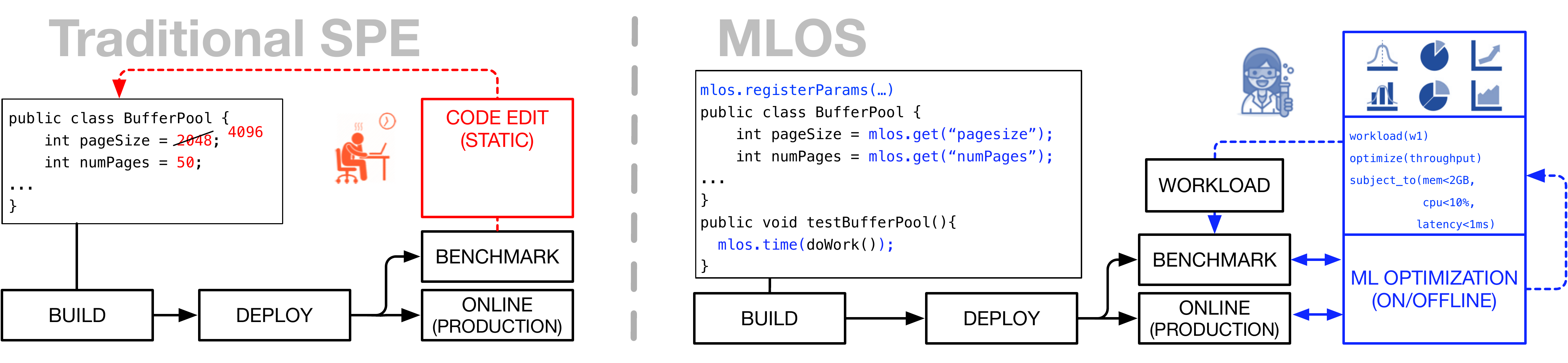}
    \vspace{-2mm}
    \caption{The MLOS Developer Experience}
    \label{fig:sf1vs2}
    \vspace{-2mm}
   \end{figure*}

As a consequence, instead of being a pervasive and continuous activity (like unit testing), SPE is relegated to elite teams with local expertise and ad-hoc knowledge and tools.
The cost of SPE can only be justified where performance is critical---Microsoft and Intel employ 100s of performance engineers that jointly tune core systems for every major hw/sw release.
The alternative is accepting sub-optimal performance.
Intel's performance engineers shared that, most system software is typically 30-40\% away from what the hardware can achieve\footnote{That is the typical boost that their specialized SPE team achieves when engaged to tune previously unoptimized software for a new hardware release.}.
This situation is exaggerated by increasingly diverse hardware available in the cloud and on the edge.

Thanks to advancements in data management and machine learning, a fundamental shift in how systems are built and tuned is possible and required.
In short, we argue in favor of \emph{melding Data Science into the practice of Software Engineering to make SPE mainstream}. Beyond the maturation of Data Science (DS), MLOS\footnote{Short for Machine Learning [Optimized|Operating] Systems.} is enabled by the move to the cloud.
Cloud operations provide us with: workload observability, increased incentives to optimize software, and an unprecedented experimentation platform. 

We present MLOS, an infrastructure and engineering practice aimed at systematizing and automating the performance tuning of a system via Data Science.
With MLOS, we empower engineers with tooling for data-driven optimizations.
This improves the end-to-end process of development and long term maintenance costs, increases documentation and reproducibility of performance tuning, and leads to more performant, efficient, well tuned systems---discussed in \S\ref{sec:developerexp}.

We observe that the SPE process closely mimics the DS process, and MLOS brings the rigor and tools of DS to it.
\eat{In a way, MLOS brings the increasing rigor and tools of DS to SPE.}
Like SPE, DS relies on experimentation, and changing parameters as often as necessary.
However, in existing systems software, many parameters are hard-coded and long build times prohibit rapid experimentation.
We believe this is a major impediment to data-driven, automated SPE.
So, one key architectural decision we make in MLOS is to separate the core system and the data-driven tuning of it.
In \S\ref{sec:architecture} we describe the infrastructure in more detail, highlighting this fundamental architectural choice.

In building MLOS, we are faced with hard challenges:
\begin{inparaenum}
\item Creating a ``natural'' development experience, enabling engineers to take advantage of advanced and rigorous DS processes.
Annotations+Automation allow for a rich DS experience with minimal effort.
\item Introducing observability/tunability without disrupting the performance of the target system. 
We achieve this using a side-agent and channel to create low-cost communications with a DS experience.
\item Global optimization across thousands of parameters is an open problem.
We cope with this by decomposing the problem (as business logic does using APIs) into microbenchmarks for sub-components and by introducing a Resource Performance Interface (RPI)---the SPE equivalent of an API.
\end{inparaenum}

We highlight the results of applying our initial MLOS implementation to tackle (1) and (2) by optimizing core data structures in SQL Server in \S\ref{sec:eval}.

At Microsoft, we are developing and using MLOS within SQL~Server engineering and are in the process of open-sourcing all of the general purpose system components \eat{ at: \url{https://github.com/microsoft/MLOS}}\cite{mlosGithub}.
Furthermore, we are collaborating with several academic institutions with the intention to create an initial class and lab.

%% file: 30_architecture.tex

\section{The MLOS Developer Experience}
\label{sec:developerexp}

MLOS aims to democratize and automate SPE.
Specifically, we leverage existing Data Science technologies, including: notebooks~\cite{ads}, model training~\cite{pedregosa2011scikit,interlandi2018machine}, tracking~\cite{zaharia2018accelerating} and deployment~\cite{bai2019}.
We concentrate our net-new development on what we believe to be a key gap that prevents system developers from using DS for SPE efficiently: the lack of observability and tunability of low level system internals.

Figure~\ref{fig:sf1vs2} highlights the difference between traditional SPE and MLOS.
For traditional SPE, the developer's ``tuning loop'' involves: hard-coding "magic" constants (or at best exposing them as start-up parameters), recompiling/redeploying (or restarting) the system, running an experiment, and analyzing the results. This process is lengthy, manual, and error-prone.

By contrast, MLOS enables externally observable and dynamically tunable constants, and allows the developer to ``drive'' and track experiments in a fashion akin to a DS experience. The loop is faster and more carefully tracked (e.g., we record all experiment conditions, including OS/HW counters). Moreover, MLOS enables automated parameter tuning, allowing for continuous and dynamic re-tuning, as well as instance-level optimization (i.e., optimize for a specific hw/sw/wl context).

Our primary concern in designing MLOS is to create a \emph{natural development experience}. The use of auto-parameters (the term we use to refer to DS-tuned system parameters) must be natural and easier than hard-coding a constant.
The first step in this journey is providing APIs and annotation mechanisms that are native to the system language.
In SQL Server (\CPP), we leverage \emph{C\#} language \emph{Attributes} to declare certain parameters as tunable, and associate metrics to critical section of the component code for observability.
As we extend MLOS to more languages, we will select mechanisms idiomatic to each language (e.g., Java annotations).

Through code-gen mechanisms (\S~\ref{sec:architecture}),
each of these parameters and their associated metrics is made observable and tunable external to the core system
without impacting the delicate performance inner-loop of the system---our second core challenge.
This externalization allows us to provide an agile notebook-driven experience (similar to a standard DS workbench). This enables a software engineer with rudimentary DS skills to visualize, analyze, model, and optimize component parameters.
We require the developer to provide (micro)benchmarks, but minimal input regarding the choice of optimization function (e.g., maximize the throughput of a component), subject to certain constraints (e.g., memory/cpu/latency caps) and the choice of workload(s).

In many cases simple AutoML or well chosen default optimization pipelines will likely provide a significant initial lift over the manual configuration.
The reason is simple: we compare a static/global configuration of parameters to a one dynamically tuned for a given hw/sw/wl.
Another value-add from MLOS is that the developer need only provide a few application level metrics specific to their component (e.g., timing of a critical section) and MLOS automatically gathers a large amount of contextual information (e.g., OS/HW counters).
These are leveraged to gain additional insight and can help in modeling a system.
Finally, in MLOS we follow DS best practices and integrate with tools such as MLFlow~\cite{zaharia2018accelerating} for versioning and tracking of all models/experiments.
On its own this provides rigor and reproducibility and makes the SPE process continuous rather than a one-off optimization that grows stale over time.
To this end we expect that in most cases the ``Data Science'' aspects of MLOS will deliver gains, more than its ML sophistication.

\textbf{Resource Performance Interface (RPI)}
A developer can optimize either the end-to-end system (e.g., TPC-E on SQL Server) or a specific component (e.g., microbenchmark on spinlocks, as we do in \S\ref{sec:eval}).
Locally optimizing individual components out of \emph{context} may not align with end-to-end optimization since those components typically share and compete for resources in the larger system.
For this reason we introduce the notion of a \emph{Resource Performance Interface (RPI)}. Conceptually an RPI allows the developer to define the \emph{context} of the component in the larger system in the form of resource and performance expectations.
Practically, this describes an acceptable ``envelope'' of resources for a given component stressed by a chosen workload.
Note that the values here may be initially specified by the developer, or learned from an existing system (e.g., during build-test runs).
RPIs provide the grounding for systematic performance regression testing, at the component level.
This is key to decompose the performance behavior of a large system, and it is the moral equivalent of an API for the non-functional performance aspects of a system.
Importantly, the RPI is \emph{not} specified as part of the system code, but as part of the DS experience.
The rationale for this is that the same piece of business logic, might be used in different parts of a system or in different environments and could be optimized differently. More precisely this means that the same API could be associated with multiple RPIs depending on the context of usage.
\\
Next, we describe how we deliver this developer experience.

\subsection{The MLOS Architecture}
\label{sec:architecture}

   \begin{figure}[t]
    \centering
    \vspace{-2mm}
    \includegraphics[width=0.9\columnwidth]{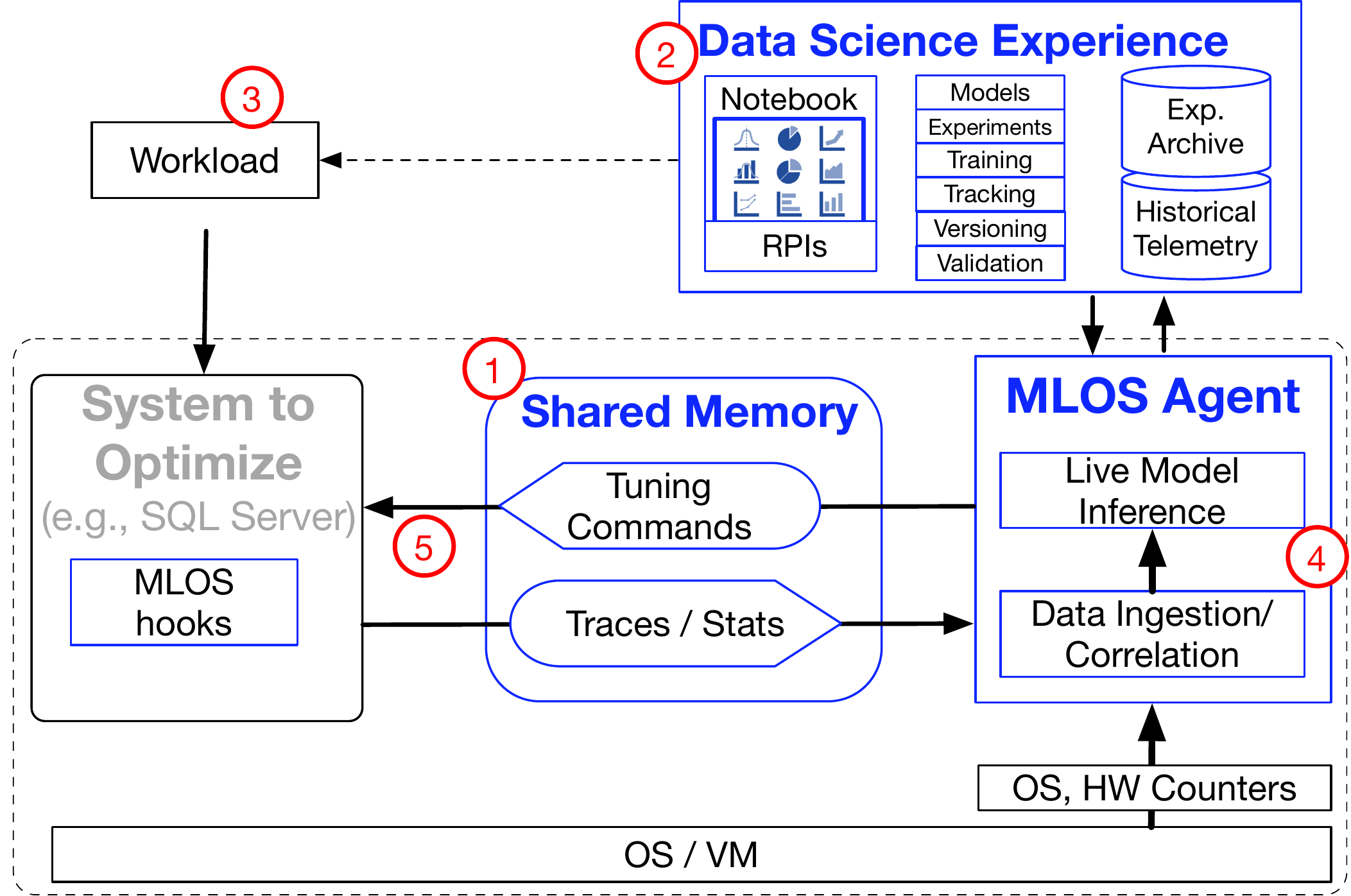}
    \vspace{-2mm}
    \caption{The MLOS Architecture}
    \vspace{-4mm}
    \label{fig:architecture}
   \end{figure}

A fundamental choice in MLOS is to enable DS-driven optimization without fundamentally modifying the underlying system.
The practical motivation comes from our initial target case: SQL Server. 
This massive codebase is hard to modify, and the size of it makes the build process very costly, slowing the experimental SPE loop.
Next we describe the mechanics of how we decouple the system from the modeling and tuning of it.
The key challenge here is to provide observability and tunability with minimal impact to the underlying system performance---a performance variant of the Socratic oath.

MLOS is depicted in Figure~\ref{fig:architecture} and operates as follows:
\\
(1) Given an annotated code base (per \S\ref{sec:developerexp}), MLOS performs code gen of:
    a) hooks in the system,
    b) a low latency shared memory communication channel, and
    c) the MLOS Agent (a daemon operating side-by-side the main system).
    This is a high-performance mechanism to externalize all tuning. 
\\
(2) A Data Science experience is enabled on top of the application metrics and OS/HW counter data collected. Developers can leverage popular data science tools for visualization, modeling, and tracking of experiments.
\\
(3) The Data Science experience also provides generic APIs to drive system-specific workloads, allowing us to focus on collecting data points of interest interactively or automatically during CI/CD and complement observations of live systems.
\\
(4) Models, Rules, and Optimizations are deployed via the DS experience into the MLOS Agent, where they are hosted for online inferencing based on live and contextual conditions (e.g., observed load on the server, internal system metrics).
\\
(5) Finally commands are sent via the shared memory channel and enacted by the MLOS hooks (e.g., updated the value for the maximum spin count of a spinlock).

 \begin{figure}[t]
    \centering
    \vspace{-2mm}
    \includegraphics[width=\columnwidth]{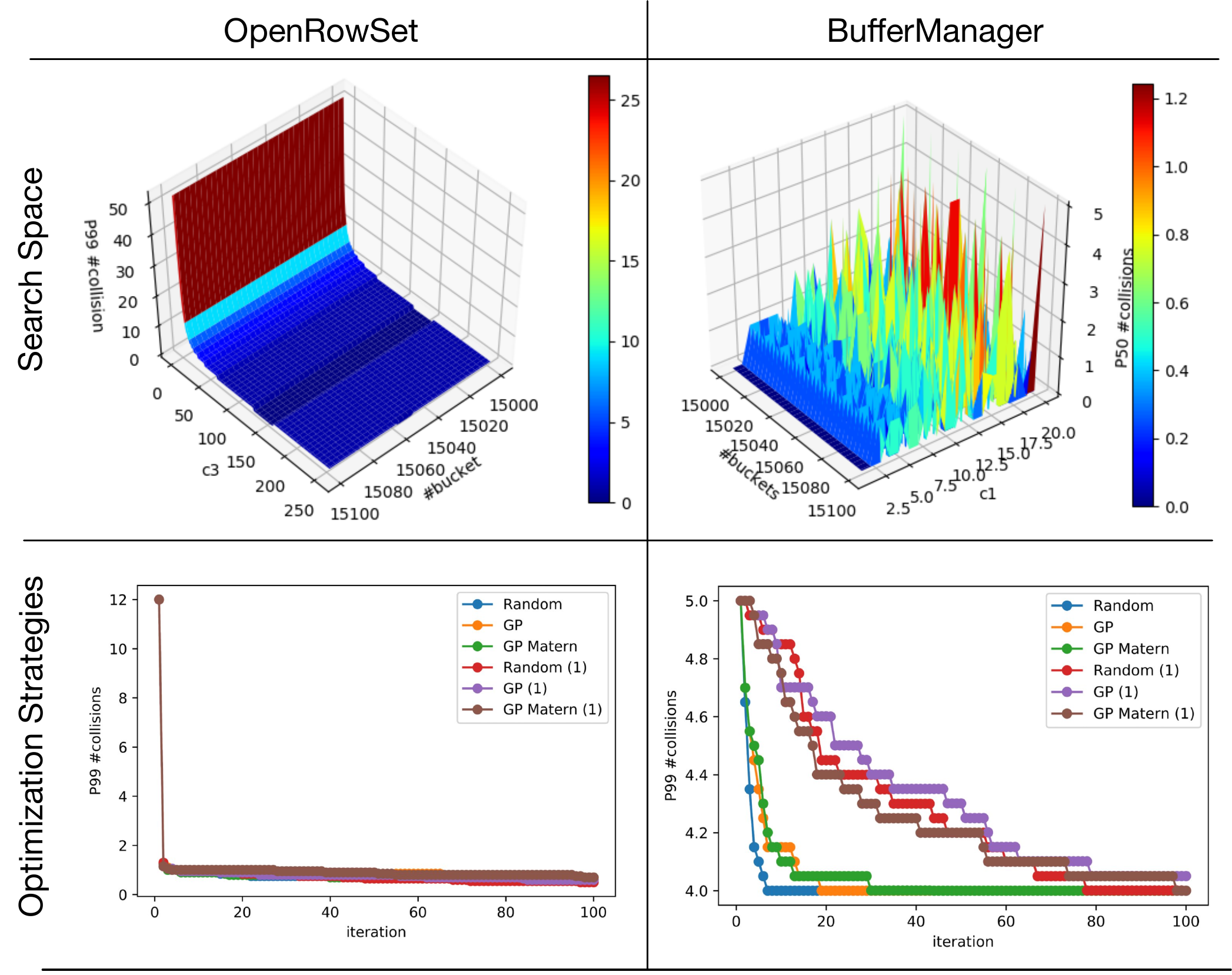}
    \vspace{-4mm}
    \caption{Tuning of two hash-tables in SQL Server}
    \vspace{-4mm}
    \label{fig:hashtable-exp}
   \end{figure}

This approach creates the independence required for fast experimentation, and reduces to a minimum the impact on the system inner-loop.
Clearly not all parameters are easily tuned dynamically, as some would incur costly re-initialization operations.
As we gain experience with MLOS in practice we will develop simple development patterns and libraries of components that are amenable to dynamic tuning.
This is why we focus our first application of MLOS to core datastructures and functions in SQL Server: SQLOS \cite{SQLOS}.

%% file: 50_evaluation.tex

\section{Evaluation}
\label{sec:eval}

Our first experiments focus on tuning the parameters of hashtables in SQL Server.
Figure~\ref{fig:hashtable-exp} shows the effects of tuning two instances (OpenRowSet and BufferManager). 
We observe: 1) DS-driven optimization can improve against SQL Server highly tuned parameters by 20\% to 90\% (the initial point in the  strategy graphs), 2) the optimization surface for the same component may vary with the workload (smooth for OpenRowSet and very jagged for BufferManager) and 3) Random Search (RS) performs very competitively with Bayesian Optimization (BO), shown using Gausian Processes (GP), GP with Mattern 3/2 kernels, and 4) optimizing multiple parameters at a time can be faster than one at a time (lines marked as (1) in graph). In other experiments (ommited due to space) BO was more efficient, indicating that the choice of optimization mechanism is non-trivial and our agile DS experience is indeed important. 

\begin{figure}[t]
    \centering
    \vspace{-2mm}
    \includegraphics[width=\columnwidth]{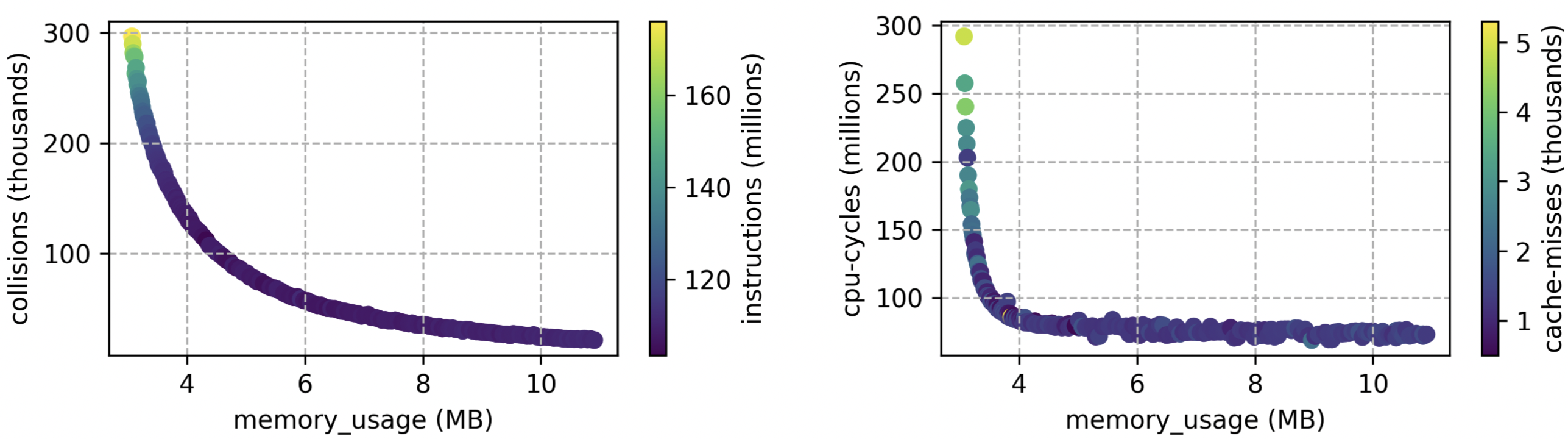}
    \vspace{-4mm}
    \caption{HW/OS Counters to tune in a resource-aware fashion}
    \vspace{-2mm}
    \label{fig:hashtable-exp2}
   \end{figure}

Next we showcase how the ability of MLOS to gather HW/OS counters during experiments allows us to adapt to different operating conditions.
Leveraging a similar Hashtable experiment Figure~\ref{fig:hashtable-exp2} shows that a larger hashtable (more buckets, hence more memory), reduces collisions (app metric) leading to better latency.
Interestingly, up to 5MB of allocation this also offers a good reduction of cpu-cycles and cache-misses (HW metrics), but this effect disappears beyond that.
This indicates that at least 5MB allocation is advisable to save CPU resources in most conditions, but that past that the tradeoff of collisions/memory dominates.
Importantly, automatic collection of HW/OS counters make it easy to observe these tradeoffs, and the MLOS optimizer allows the developer to focus on the desired objective (e.g., a tradeoff of resources and performance) abstracting away the parameter fiddling.
Finally, we note that MLOS can perform this tuning continously adapting to changing hw/sw/wl conditions.

\begin{figure}[t]
    \centering
    \vspace{-2mm}
    \includegraphics[width=0.8\columnwidth]{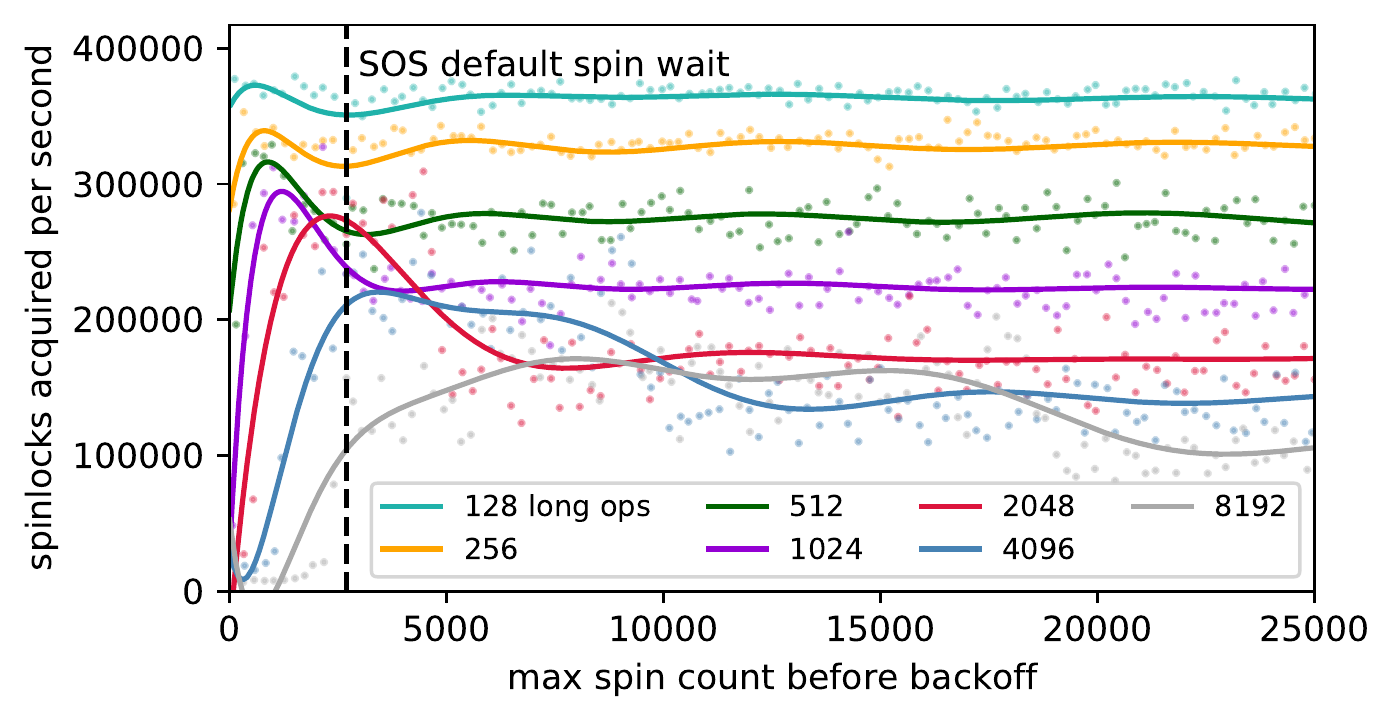}
    \vspace{-4mm}
    \caption{Optimal spinlock polling length varies by workload.}  
    \label{fig:spinlock-tuning-by-workload-length}
    \vspace{-3mm}
   \end{figure}

Our last experiment showcases how the workload may influence optimal parameters choice. Figure~\ref{fig:spinlock-tuning-by-workload-length} shows the performance of a spinlock when changing the maximum spinning before backoff. We show 7 workloads each comprised of several threads performing little work, and one performing a larger and larger number of operations under the lock. Subtle changes in the workload (or hardware platform) can substantially affect the optimal choice of parameters. 

\textbf{Discussion} These experiments show great promise for component-level optimizations. A key open challenge: how to tackle the impractical parameter space for global optimization. Our current practical solution is to leverage the power of RPIs combined with developer intuition. More research in this space is needed. 

%% file: 60_related.tex

\section{Related Work}

MLOS and Software~2.0~\cite{software20} are related, but not the same.
MLOS does not make a distinction between developers who write code ("Developer 1.0") and those that curate data and train models ("Developer 2.0").
Instead, MLOS augments systems developers with enough data science tools and methodology to meet the challenges of the day.
Similarly, MLOS isn't geared at problems that can only be solved by replacing functions with models (e.g. image recognition).
Instead, it fuses traditional solutions (e.g. OS/HW measurement of benchmarks) to problems solvable with them (e.g. hashtable optimization) with data science to organize and optimize them for the context in which they are applied.
Recent work \cite{kraska2018case, idreos2018data, marcus2019neo, ding2019alex} on replacing sub-components of a DBMS with models is incredibly promising, but as of the time of this writing still brittle for our production applications.  

Another key area of related work, is tools that tune external parameters of a system \cite{duan2009tuning, van2017automatic, pavlo2019external, zhu2017bestconfig}.
Closer points of comparison are approaches such as \cite{marcus2020bao,wu2018towards,siddiqui2020cost,jindal2019peregrine} that leverage ML to guide/improve existing query optimizers.  All of these efforts are complementary to MLOS, which focuses on making system internals observable and tunable. We are in fact, considering to port some of them onto the MLOS infrastructure.

Past efforts have also tried to address this problem at a library \cite{carbune2018smartchoices, eastep2011smart} or compiler level \cite{fursin2011milepost, ashouri2018survey, chen2016autofdo}.
In some ways these are early applications of MLOS style optimization for those areas.
While complementary, though don't focus on exposing the DS for the developer to improve with as well.

%% file: 70_conclusion.tex

\section{Conclusion}
Software Performance Engineering (SPE) is a very valuable but costly activity. With MLOS we fill an infrastructure gap that will enable developers to leverage the tools and practices of Data Science (DS).